\documentclass{elsart}

\usepackage[square,comma]{natbib}
\usepackage{graphicx}
\usepackage{pxfonts}
\usepackage{lineno}

\usepackage{amssymb}

\journal{}

\begin{document}

\thispagestyle{empty}
\begin{Large}
\textbf{DEUTSCHES ELEKTRONEN-SYNCHROTRON}

\textbf{\large{Ein Forschungszentrum der
Helmholtz-Gemeinschaft}\\}
\end{Large}

DESY 10-006

January 2010

\begin{eqnarray}
\nonumber &&\cr \nonumber && \cr \nonumber &&\cr
\end{eqnarray}
\begin{eqnarray}
\nonumber
\end{eqnarray}
\begin{center}
\begin{Large}
\textbf{Scheme for simultaneous generation of three-color  ten
GW-level X-ray pulses from baseline XFEL undulator and multi-user
distribution system for XFEL laboratory}
\end{Large}
\begin{eqnarray}
\nonumber &&\cr \nonumber && \cr
\end{eqnarray}

\begin{large}
Gianluca Geloni,
\end{large}
\textsl{\\European XFEL GmbH, Hamburg}
\begin{large}

Vitali Kocharyan and Evgeni Saldin
\end{large}
\textsl{\\Deutsches Elektronen-Synchrotron DESY, Hamburg}
\begin{eqnarray}
\nonumber
\end{eqnarray}
\begin{eqnarray}
\nonumber
\end{eqnarray}
ISSN 0418-9833
\begin{eqnarray}
\nonumber
\end{eqnarray}
\begin{large}
\textbf{NOTKESTRASSE 85 - 22607 HAMBURG}
\end{large}
\end{center}
\clearpage
\newpage

\begin{frontmatter}



\title{Scheme for simultaneous generation of three-color  ten GW-level X-ray
pulses from baseline XFEL undulator and multi-user distribution
system for XFEL laboratory}


\author[XFEL]{Gianluca Geloni\thanksref{corr},}
\thanks[corr]{Corresponding Author. E-mail address: gianluca.geloni@xfel.eu}
\author[DESY]{Vitali Kocharyan}
\author[DESY]{and Evgeni Saldin}

\address[XFEL]{European XFEL GmbH, Hamburg, Germany}
\address[DESY]{Deutsches Elektronen-Synchrotron (DESY), Hamburg,
Germany}

\begin{abstract}
The baseline design of present XFEL projects only considers the
production of a single photon beam at fixed wavelength from each
baseline undulator. At variance, the scheme described in this
paper considers the simultaneous production of high intensity SASE
FEL radiation at three different wavelengths.  We present a
feasibility study of our scheme, and we make exemplifications with
parameters of the baseline SASE2 line of the European XFEL
operating in simultaneous mode at $0.05$ nm, $0.15$ nm and $0.4$
nm. Our technique for generating the two colors at $0.05$ nm and
$0.15$ nm is based in essence on a "fresh bunch" technique. For
the generation of radiation at 0.4 nm we propose to use an
"afterburner" technique. Implementation of these techniques does
not perturb the baseline mode of operation of the SASE2 undulator.
The present paper also describes an efficient way to obtain a
multi-user facility. It is shown that, although the XFEL photon
beam from a given undulator is meant for a single user, movable
multilayer X-ray mirrors can be used to serve many users
simultaneously. The proposed photon beam distribution system would
allow to switch the FEL beam quickly between many experiments in
order to make an efficient use of the source. Distribution of
photons is achieved on the basis of pulse trains and it is
possible to distribute the multicolor photon beam among many
independent beam lines, thereby enabling many users to work in
parallel with different wavelengths.
\end{abstract}

%
%
%
\end{frontmatter}



\section{\label{sec:intro} Introduction}

Two recent papers of us \cite{OUR01} and \cite{OUR02} discuss the
exploitation of the high-performance beam-formation systems
available for XFEL facilities. Recent results at LCLS
\cite{LCLS2,DING} demonstrate that they can work as in the ideal
operation scenario. In particular, at LCLS, the small
electron-beam emittance achieved ($0.4 \mu$m with $0.25$ nC
charge) allows saturation within $20$ undulator cells, out of the
$33$ available.

This optimal scenario should be exploited to provide users with
the best possible fruition opportunities. Based on the fresh bunch
technique \cite{HUAYU}-\cite{SAL2},  we suggested a method
\cite{OUR01} to use the extra-undulator length available to
provide two short (sub-ten fs), powerful (ten GW-level) pulses of
coherent x-ray radiation at different wavelengths \footnote{Within
the full range of tunability of SASE2, i.e. from $0.1$ nm to $0.4$
nm.} for pump-probe experiments at XFELs, with minimal hardware
changes to the baseline setup. In the following paper \cite{OUR02}
we considered a similar approach for extending the spectral range
accessible to the European XFEL down to $0.05$ nm, by creating
simultaneously two ten-fs, ten-GW pulses at $0.15$ nm and $0.05$
nm. We presented feasibility studies and we made exemplifications
with the parameters of the SASE2 line of the European XFEL. In the
present paper we extend the method in \cite{OUR02} to include
simultaneous operation at three colors, $0.05$ nm, $0.15$ nm and
$0.4$ nm, and we subsequently describe a multi-user distribution
system based on multilayers, which is proposed as an efficient way
to distribute radiation to many users.

\begin{figure}
\begin{center}
\includegraphics*[width=100mm]{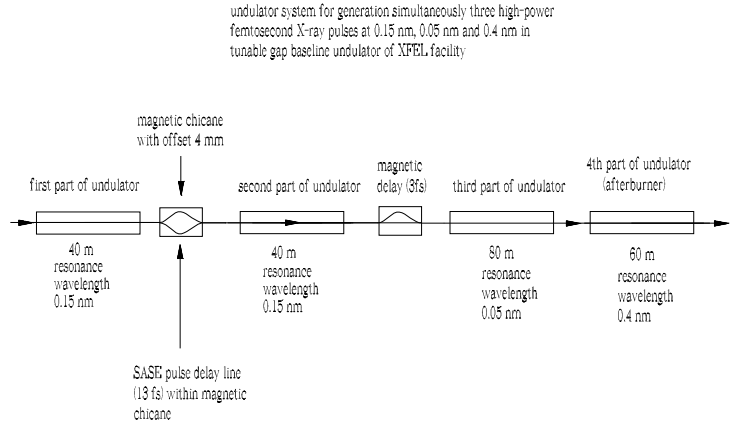}
\caption{\label{multi2}Design of undulator system for high-power
three-color femtosecond X-ray pulse generation. }
\end{center}
\end{figure}

\begin{table}
\caption{Parameters for the short pulse mode used in this paper.
The undulator parameters are the same of those for the European
XFEL, SASE2, at 17.5 GeV electron energy.}

\begin{small}\begin{tabular}{ l c c}
\hline
& ~ Units &  Short pulse mode \\
\hline
Undulator period      & mm                  & 47.9   \\
Undulator length      & m                   & 256.2  \\
Segment length        & m                   & 6.1    \\
Number of segments    & -                   & 42     \\
K parameter (rms)     & -                   & 2.513-4.300  \\
$\beta$               & m                   & 17     \\
Wavelength            & nm                  & 0.15 - 4.0   \\
Energy                & GeV                 & 17.5   \\
Charge                & nC                  & 0.025  \\
Bunch length (rms)    & $\mu$m              & 1.0    \\
Normalized emittance  & mm~mrad             & 0.4    \\
Energy spread         & MeV                 & 1.5    \\
\hline
\end{tabular}\end{small}
\label{tab:fel-par}
\end{table}
The extension of the method proposed in \cite{OUR02}, described in
section \ref{due}, is sketched in Fig. \ref{multi2}, and is
justified by the length of the SASE2 undulator (see Table
\ref{tab:fel-par}), which is about a hundred meters longer than
what is needed to implement the technique \cite{OUR02}. The basic
idea is to exploit this extra-available length in the
"afterburner" mode by letting the spent electron beam emit SASE
radiation at a longer wavelength, $0.4$ nm, for which the electron
beam quality is still good enough to lase. In this way, three
superimposed high-intensity pulses of radiation are produced.

\begin{figure}
\begin{center}
\includegraphics*[width=100mm]{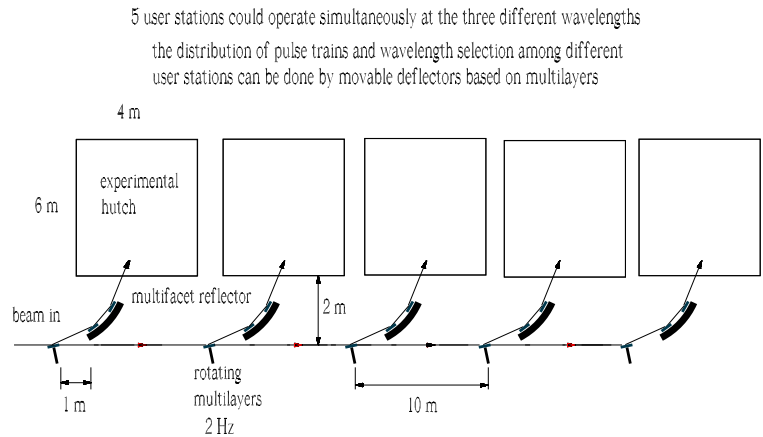}
\caption{\label{beamline} Proposed SASE undulator beam line. A
photon beam distribution system based on movable multilayer X-ray
mirrors can provide an efficient way to obtain a multi-user
facility. Distribution of photons is achieved on the basis of
pulse trains and it is possible to serve simultaneously 5 user
stations with (train) repetition rate 2 Hz at three different
wavelengths.  }
\end{center}
\end{figure}
Finally, we speculate to optimize the fruition of the three color
pulses by installing a photon beam distribution system based on
multilayer X-ray mirrors, as sketched in Fig. \ref{beamline} and
discussed in section \ref{tre} \cite{SALT}. With the help of this
technique, many user stations can be fed at the same time with the
same high-quality photon beam.

\section{Feasibility study \label{due}}

\begin{figure}[tb]
\includegraphics[width=1.0\textwidth]{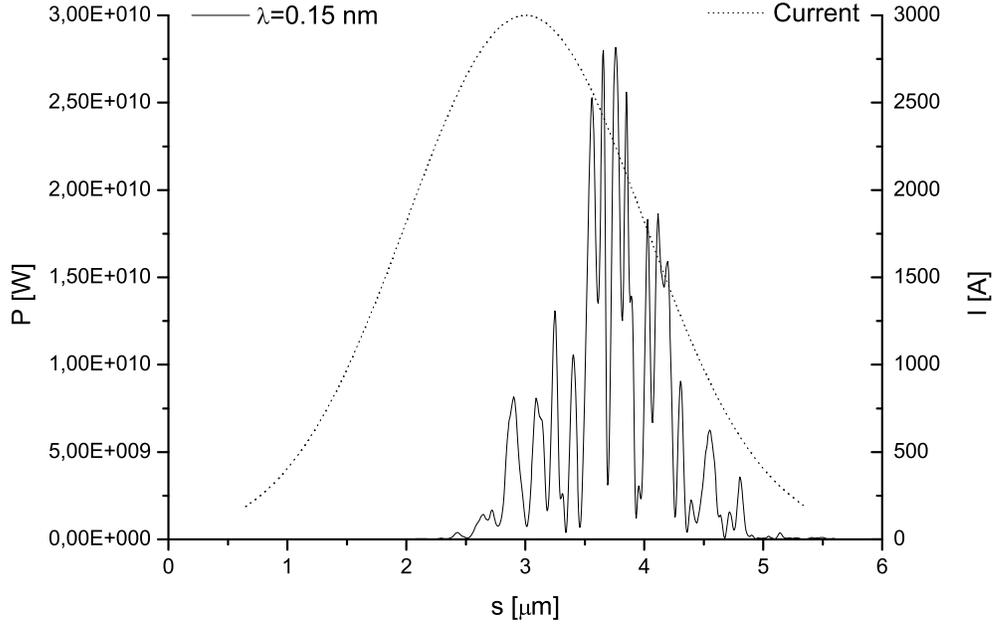}
\caption{Beam power distribution at the entrance of the third
stage, after the magnetic delay, at $0.15$ nm.} \label{firstsh}
\end{figure}

\begin{figure}[tb]
\includegraphics[width=1.0\textwidth]{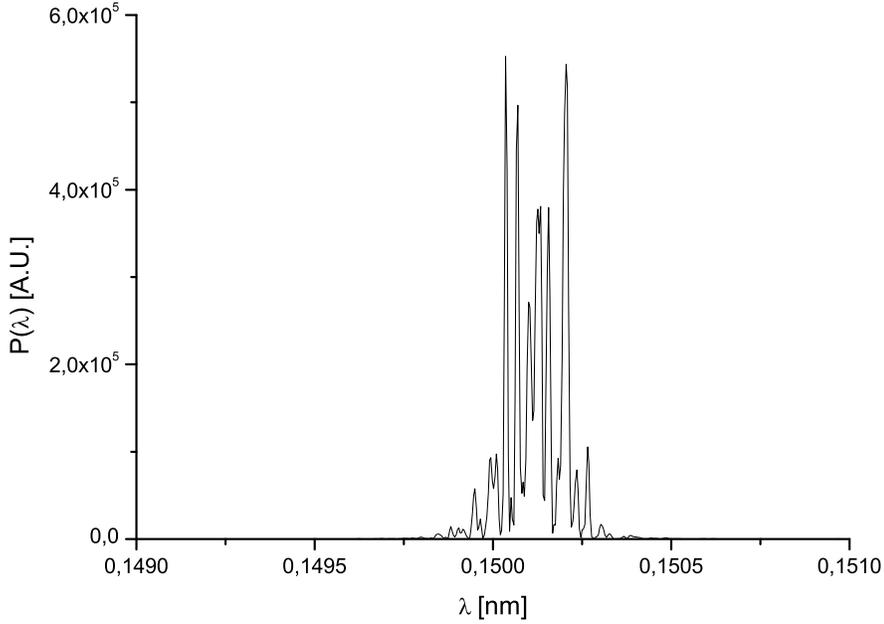}
\caption{First harmonic beam power spectrum at the entrance of the
third stage.} \label{IIstageS}
\end{figure}

\begin{figure}[tb]
\includegraphics[width=1.0\textwidth]{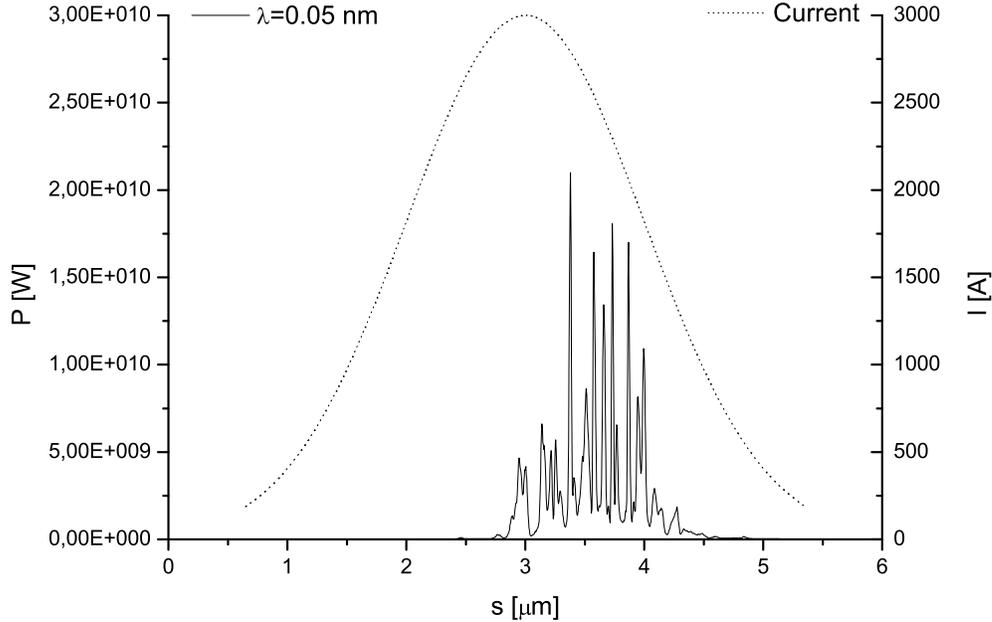}
\caption{Beam power distribution at $0.05$ nm at the end of the
third stage, which is $11$ cells-long ($67.1$ m).}
\label{IIIstageP}
\end{figure}

\begin{figure}[tb]
\includegraphics[width=1.0\textwidth]{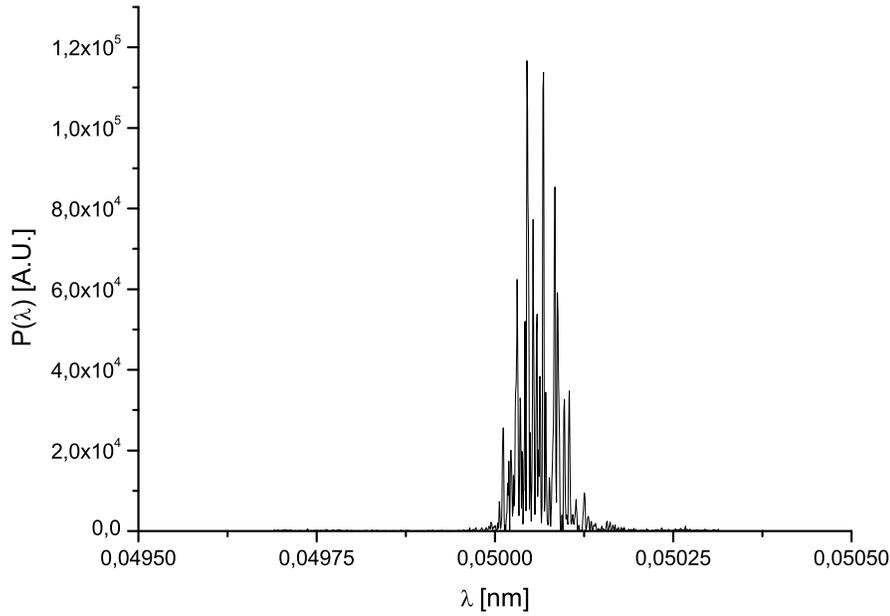}
\caption{Third harmonic beam power spectrum at the end of the
third stage.} \label{IIIstageS}
\end{figure}

The starting point for the feasibility study of the three-color
technique is the arrival point of reference \cite{OUR02}, where we
simulated the simultaneous production of two ten-fs, ten-GW pulses
at $0.15$ nm and $0.05$ nm. We sum up these results in Fig.
\ref{firstsh}-\ref{IIstageS} and Fig.
\ref{IIIstageP}-\ref{IIIstageS}, which show respectively the beam
power and spectrum distribution at $0.15$ nm at the entrance of
the third stage, and the beam power and spectrum distribution at
$0.05$ nm at the end of the third stage of Fig. \ref{multi2}.



\begin{figure}[tb]
\includegraphics[width=0.5\textwidth]{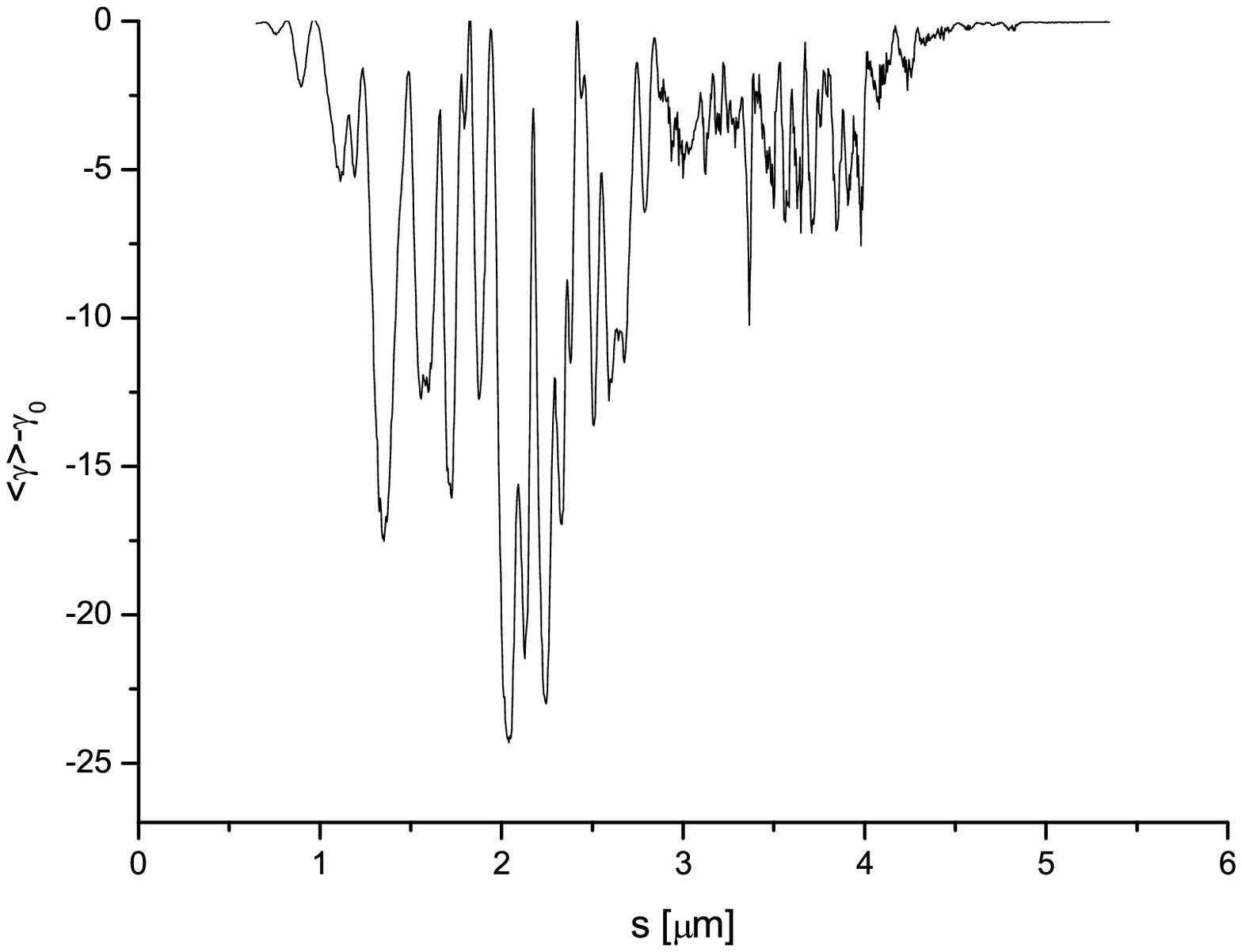}
\includegraphics[width=0.5\textwidth]{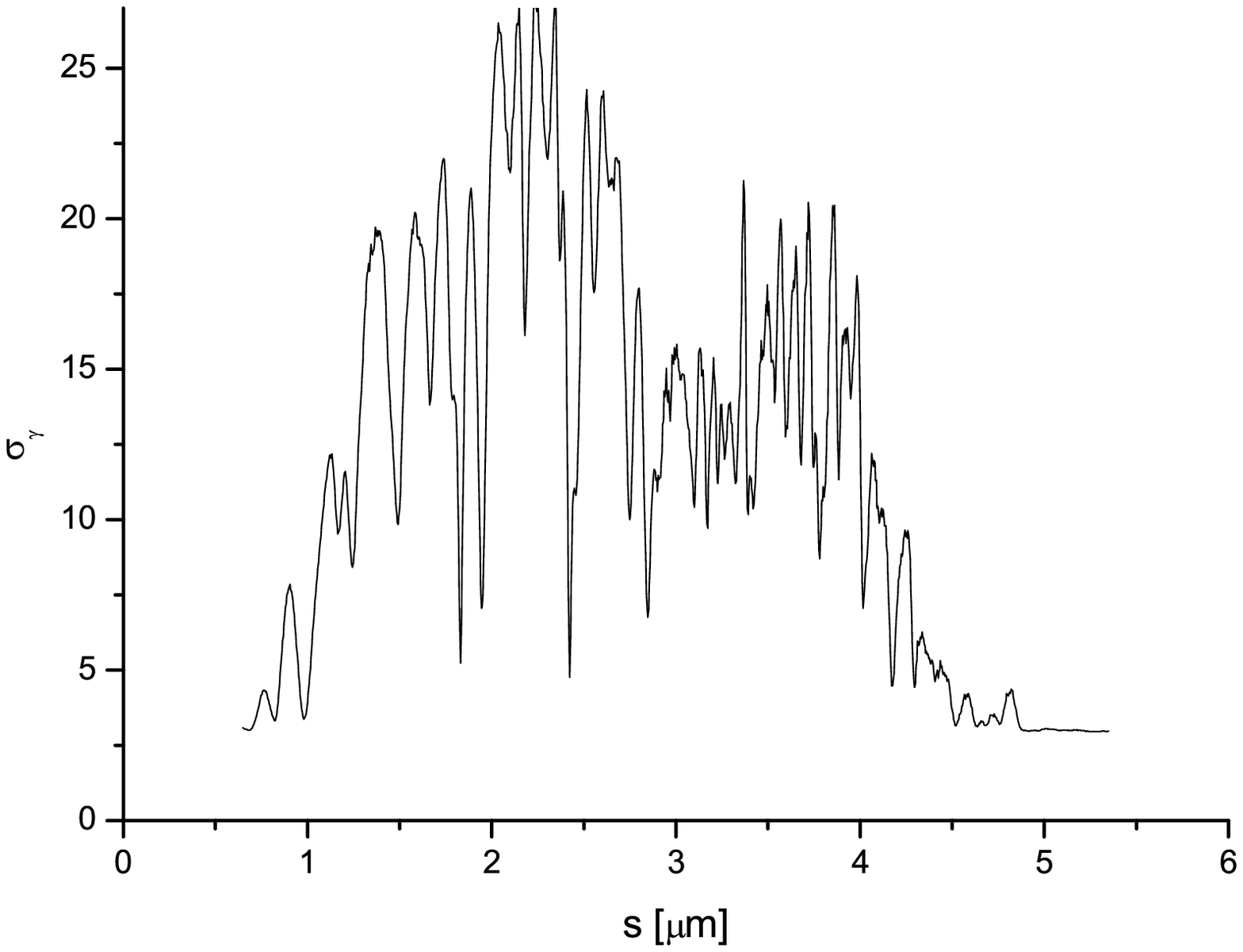}
\caption{Electron beam energy loss (left) and induced energy
spread (right) at the end of the third stage, which is $11$
cells-long ($67.1$ m).} \label{IIIstageen}
\end{figure}
At the entrance of the fourth stage of Fig, \ref{multi2}, the
electron beam is spent, and its quality is deteriorated by the
lasing process. This fact can be seen by inspecting Fig.
\ref{IIIstageen}, where we show the induced energy spread and the
energy loss accumulated from the previous FEL interactions.
Therefore it is not possible to lase at the same short wavelengths
as in the previous stages. However, requirements on the beam
quality for lasing are relaxed going to longer wavelength. This is
essentially the "afterburner" concept \cite{AFTE}. Using as before
Genesis 1.3 \cite{GENE} we generate an electron bunch with energy
loss and energy spread distributions as in Fig. \ref{IIIstageen},
and we simulate the SASE process in the fourth stage of Fig.
\ref{multi2}.

\begin{figure}[tb]
\includegraphics[width=1.0\textwidth]{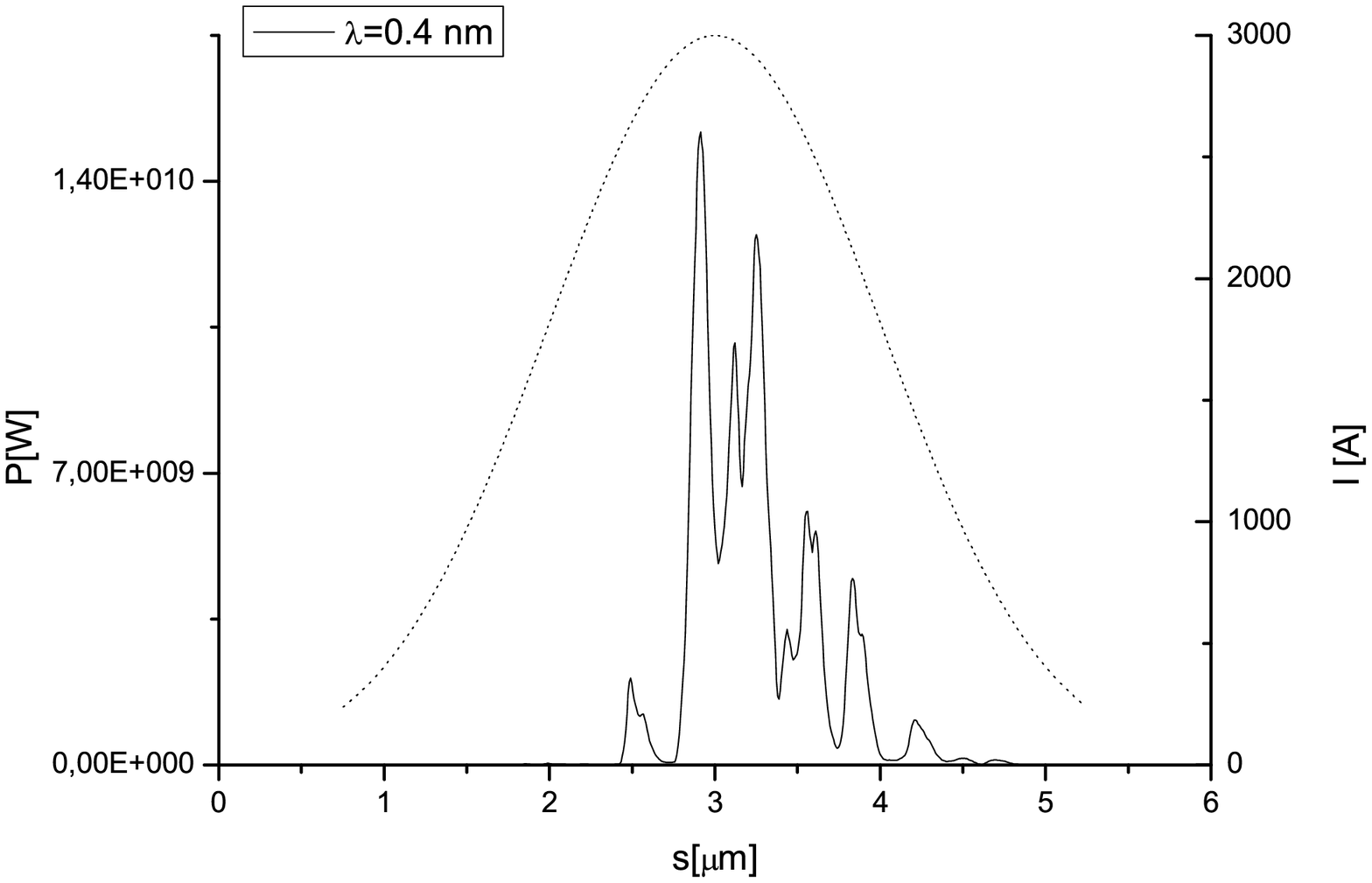}
\caption{Beam power distribution at $0.4$ nm at the end of the
fourth stage,  which is $9$ cells-long ($54.9$ m).}
\label{IVstageP}
\end{figure}
\begin{figure}[tb]
\includegraphics[width=1.0\textwidth]{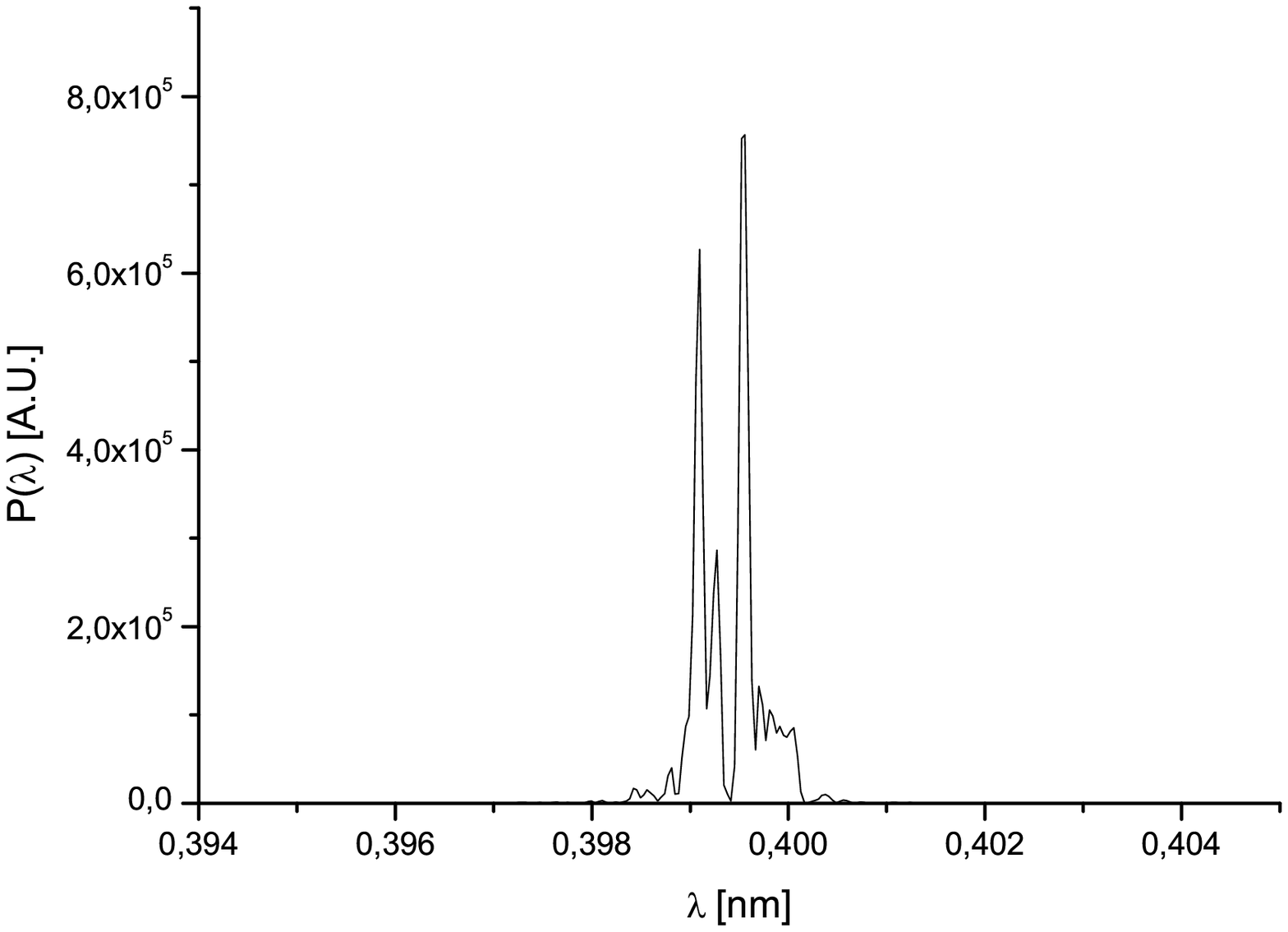}
\caption{Beam power spectrum around $0.4$ nm, at the end of the
fourth stage.} \label{IVstageS}
\end{figure}
The result of this simulation is shown in Fig. \ref{IVstageP} and
Fig. \ref{IVstageS} in terms of power and spectrum distribution
respectively. We stop the simulation after 9 cells, as we obtain a
power level comparable with those in Fig. \ref{firstsh} and Fig.
\ref{IIIstageP}. However, extra undulator length is available, and
the output power of the third color may in principle be increased
far beyond $10$ GW.

\section{Multi-user distribution system based on multilayers. \label{tre}}

The typical layout of a SASE FEL is a linear arrangement in which
the injector, accelerator, bunch compressors and undulator are
nearly collinear, and in which the electron beam does not change
the direction between accelerator and undulators. However, it is
desirable that an X-ray FEL laboratory could serve tens of
experimental stations which should operate independently according
to the needs of the user community. In this section we describe a
beam distribution system, sketched in Fig. \ref{beamline}, which
may allow to switch the FEL beam quickly between many experiments
in order to make an efficient use of the source. Many applications
require only very high peak brilliance. Experiments for which the
average brilliance is not critical could operate simultaneously at
the three different radiation wavelengths available.

\begin{figure}
\begin{center}
\includegraphics*[width=100mm]{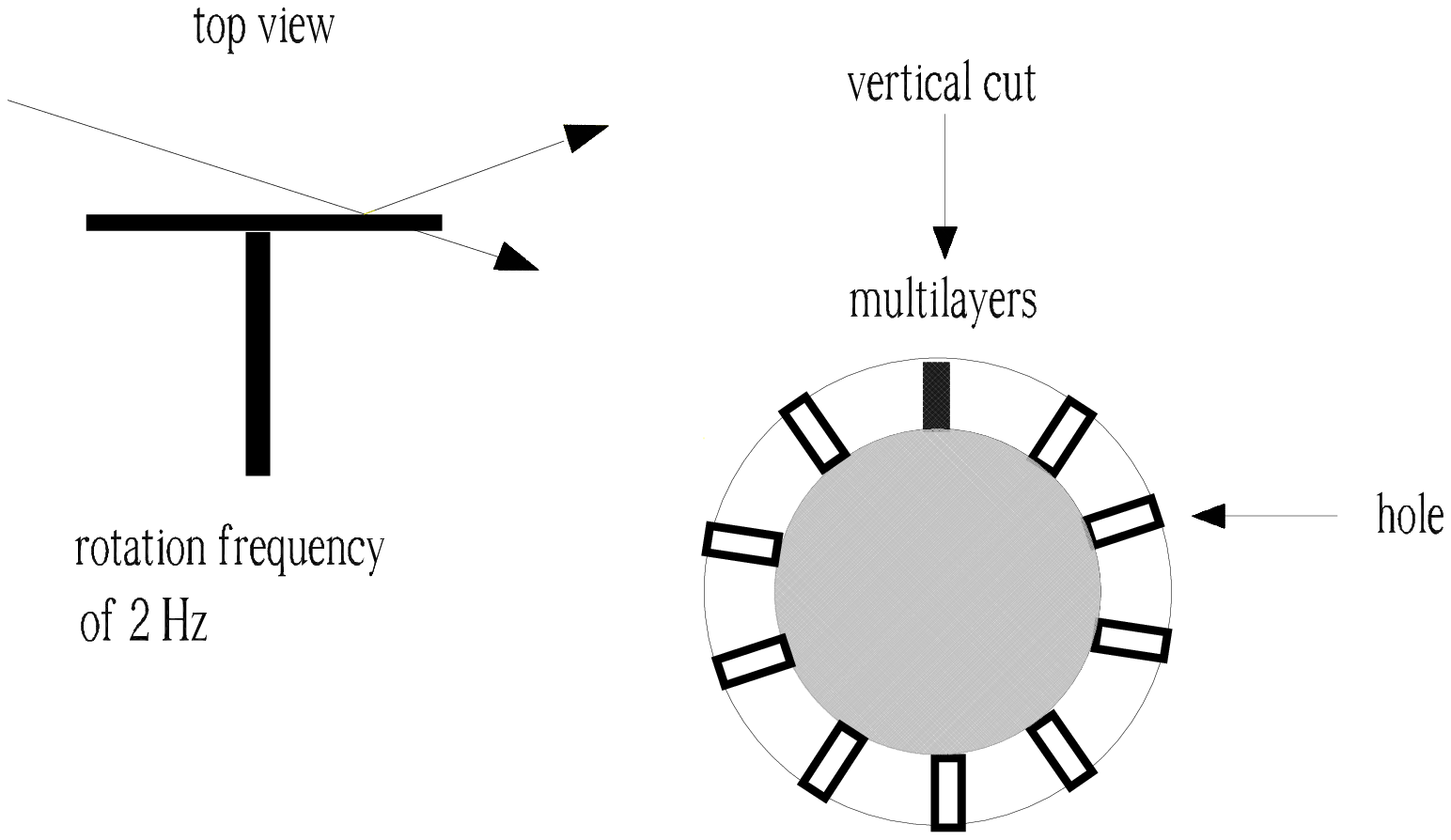}
\caption{\label{multi} Different views of a switching multilayer
mirror.  }
\end{center}
\end{figure}
\begin{figure}
\begin{center}
\includegraphics*[width=100mm]{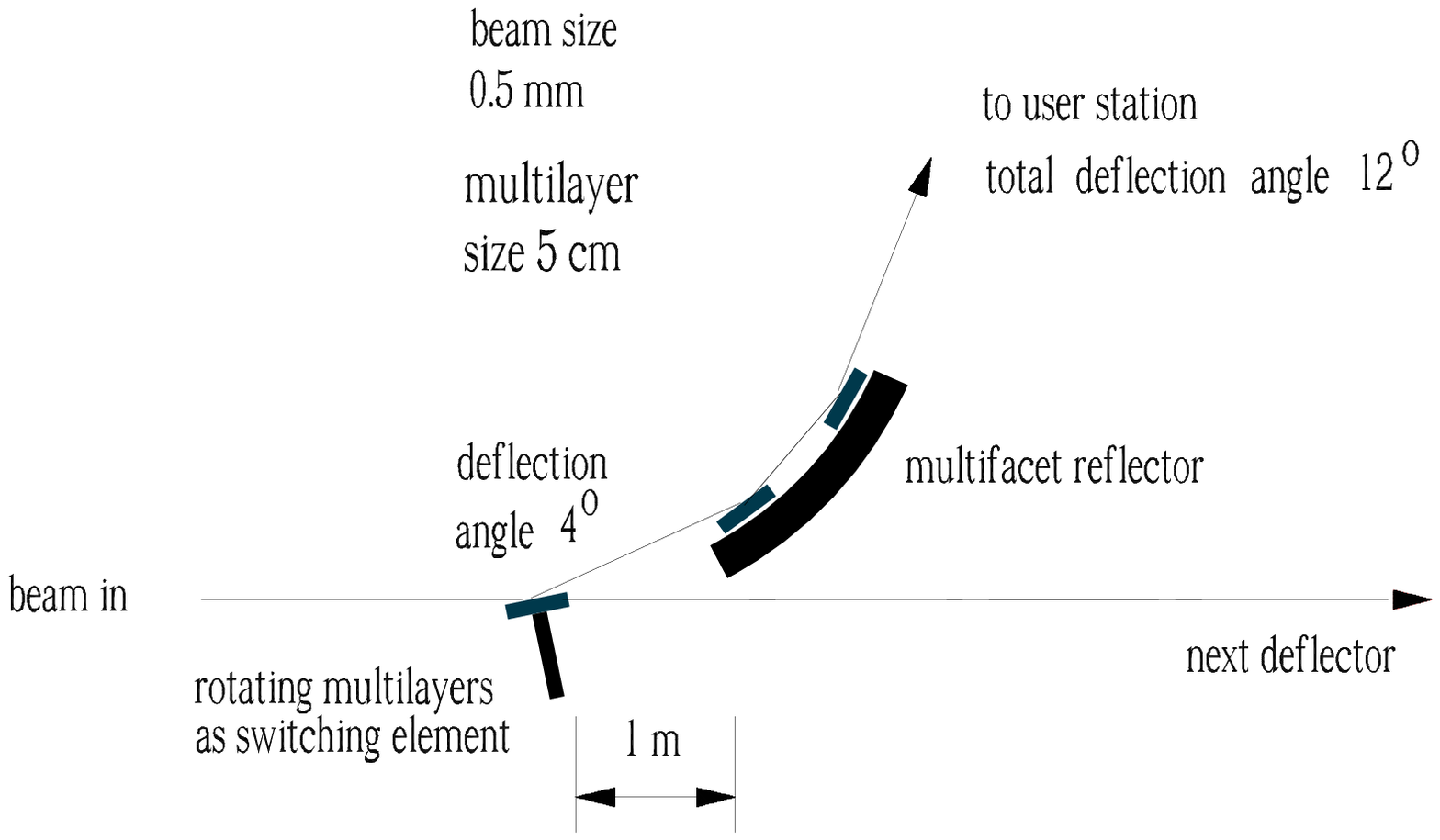}
\caption{\label{deflector} Concept of the photon beam deflector
based on multilayer X-ray mirror.  }
\end{center}
\end{figure}

The technical approach adopted in this variant of the XFEL
laboratory design makes use of movable multilayer X-ray mirrors.
Layered structures with usually composed by two alternating
materials with low and high density respectively. These structures
play an important role in synchrotron X-ray optics
\cite{WINDT}-\cite{MORA}. Typical multilayers used as optical
elements at third generation synchrotrons provide a spectral
bandwidth of $1 \%$ to $5 \%$. Typical glancing angles are of the
order of a degree and thus lie between the mrad-wide angles of
X-ray mirrors and the much wider $10$ degree-wide Bragg angles of
single crystals. The angular acceptance of X-ray multilayer
mirrors is of the order of a mrad (for a bandwidth $1\%$). As a
rule, from 100 to 400 layers (each about $10$ nm-thick)
participate in the effective reflection in such mirrors. About $90
\%$ peak reflectivity was achieved for wavelengths around 0.1 nm.
Computer simulations are in very good agreement with experimental
results in all cases, so that efficiencies can be safely
predicted.

Fig. \ref{multi}   shows the relevant photon beamline
configuration. The distribution of pulse trains among the
different user stations can be done by movable deflectors. A
schematic diagram of a movable deflector is shown in Fig.
\ref{deflector}. Its key components include rotating multilayers
and a multifacet reflector.

The advantages of using multilayer mirrors as movable photon beam
deflectors are manyfold. First, multilayer mirrors are
characterized by larger deflection angle compared to X-ray
mirrors. As a result, the length of a multilayer mirror required
for deflecting photon beam with transverse size order of fraction
of mm is in the ten-cm scale. In contrast to this, X-ray mirrors
performing the same function are in the meter-scale length.
Second, the bandwidth of the multilayer reflectivity is much wider
compared to crystals, and it can be possible to deflect the full
spectrum of the SASE radiation pulse without perturbation. It
should be noted that in the ideal case, multilayer mirrors keep
the angular beam divergence constant.  Third, the bandwidth of
multilayer mirrors is much smaller than unity and there is a
possibility in our case to deflect only a single color beam when
we deal with multi-color pulse generation.

In order to achieve stable photon beam deflection the alignment
accuracy of multilayer deflectors must be less than 0.1 mrad. It
is believed that technology  will enable rotating multilayers to
satisfy these requirements. The initial photon beam will thus be
transformed into $5$ different beams. The switching mirrors need
to rotate (for European XFEL laboratory case) at a frequency of 2
Hz such that each user actually receives two trains of pulses with
a duration of 1 ms per second. Note that even if the photon beam
is distributed among many users, the peak flux per user remains
untouched (apart from the losses in the deflector system). All
users will receive a photon beam of identical, high quality.

It should be noted that the deflection process happens only once
during the pass of a given photon beam through the deflector unit.
Also, the deflection process requires three multilayers mirror
only, Fig. \ref{deflector}, so that the problem of absorption of
the radiation in the distribution system does not exist. Another
advantage of this method comes from the small (one to few
microradians) angular divergence of the XFEL radiation. In fact,
in an extended sequence of deflectors (say a few tens of meters),
Fig. \ref{beamline}, the photon beam spot size remains unchanged.
Finally, another attractive feature of the proposed photon
distribution system is a high degree of flexibility: if some user
will request a full photon flux for a given time, such request may
be granted simply by "freezing" the motion of the multilayer
mirrors for a certain time, so that the full photon flux will be
directed to a single, dedicated user station.

\section{Conclusions}

In this article we discuss the possibility of extending the
technique in \cite{OUR02} to three colors. This is feasible due to
the relatively short undulator length needed for the exploitation
of the two-colors technique. The technique itself is fairly
straightforward, and consists in the use of the remaining part of
the undulator as an afterburner. We exemplified the proposed
technique for the baseline parameters of SASE2 at the European
XFEL. In particular, we showed how three sub-ten fs, 10 GW power
level pulses can be simultaneously produced at $0.4$ nm, $0.15$ nm
and $0.05$ nm. As for the method described in \cite{OUR02}, the
present method requires very limited hardware too and is low cost.
Moreover, it carries no risks for the operation of the machine in
the baseline mode.

Subsequently, we speculate on how many users may simultaneously
take advantage of these pulses in an effective way. In fact, the
typical geometry of a SASE FEL where injector, accelerator, bunch
compressors and undulator are nearly collinear, does not seem
suitable for simultaneous operation of many users. However, a
different concept of the XFEL laboratory allows one to
simultaneously serve the need of multiple users. We propose to
consider a multi-user distribution system based on multilayer
X-ray mirrors, which allow many user stations to work in parallel
with the same high-quality X-ray beams.

Finally, it should be noted that even if we discuss the case of
the European XFEL, our technique may be taken advantage of by
other facilities as well.

\section{Acknowledgements}

We are grateful to Massimo Altarelli, Reinhard Brinkmann, Serguei
Molodtsov and Edgar Weckert for their support and their interest
during the compilation of this work.


\begin{thebibliography}{99}

\bibitem{OUR01} G. Geloni, V. Kocharyan and E.~Saldin, "Scheme for femtosecond-resolution pump-probe
experiments at XFELs with two-color ten GW-level X-ray pulses",
DESY 10-004 (2010).

\bibitem{OUR02} G. Geloni, V. Kocharyan and E.~Saldin, "The potential for extending the spectral range
accessible to the European XFEL down to 0.05 nm", DESY 10-005
(2010).

\bibitem{LCLS2} P. Emma, First lasing of the LCLS X-ray FEL at 1.5 Å, in
Proceedings of PAC09, Vancouver, to be published in
http://accelconf.web.cern.ch/AccelConf/~.

\bibitem{DING} Y. Ding et al., Phys. Rev. Lett. 102, 254801
(2009).

\bibitem{HUAYU}  I. Ben-Zvi and L.H. Yu, Nucl. Instr. and Meth. A
393, 96 (1997).

\bibitem{SAL1} E. Saldin, E. Schneidmiller and M. Yurkov, Opt.
Commun. 212, 377 (2002).

\bibitem{SAL2} E. Saldin, E. Schneidmiller and M. Yurkov, Opt.
Commun., 239, 161 (2004).

\bibitem{SALT} E. Saldin, E. Schneidmiller and M. Yurkov,
TESLA-FEL 2004-02 (2004).

\bibitem{WINDT} D. Windt, Appl. Phys. Lett. 74, 2890 (1999).

\bibitem{MAJK}  E. Majkova et al., "Nanometer-scale period multilayers: thermal
stability study", report at the 7th International Conference on
the Physics of X-ray Multilayer Structures, Sapporo, Japan, 2004

\bibitem{DESC} P. Deschamps et al., J. Synchrotron Rad. 2, 14
(1995).

\bibitem{MORA} Ch. Morawe et al., SPIE Proc. 4145, 61 (2000).

\bibitem{AFTE} R. Bonifacio et al., NIM A 296, 787 (1990).
\bibitem{GENE} S Reiche et al., Nucl. Instr. and Meth. A 429, 243 (1999).




\end{thebibliography}
\end{document}